\documentclass[aps,prb,twocolumn,superscriptaddress,showpacs]{revtex4}
\usepackage{amsmath}
\usepackage{amssymb}
\usepackage{graphicx}
\usepackage{subfigure}
\usepackage{float}
\begin{document}
\title{Scaling analysis of the quasiparticle tunneling in the ${\mathbb Z}_k$ parafermion states }

\author{Qi Li}
\affiliation{Department of Physics, Chongqing University,Chongqing,401331,P.R. China}
\author{Na Jiang}
\affiliation{Department of Physics, Chongqing University,Chongqing,401331,P.R. China}
\author{Xin Wan }
\affiliation{Zhejiang Institute of Modern Physics, Zhejiang University, Hangzhou 310027,P.R. China}
\affiliation{Collaborative Innovation Center of Advanced Microstructures, Nanjing 210093,P.R. China}
\author{Zi-Xiang Hu}
\email{zxhu@cqu.edu.cn}
\affiliation{Department of Physics, Chongqing University,Chongqing,401331,P.R. China}

\date{\today}
\begin{abstract}
Quasiparticle tunneling between two counter-propagating edges through point contacts could provide information on the 
statistics of the quasiparticles.  Previous study on a disk  found a scaling behavior 
by varying the tunneling distance. It was found that in the limit with zero tunneling distance,  the Abelian quasiparticles tunneling obey the scaling analysis while the 
non-Abelian quasiparticles exhibit some non-trivial behaviors on the scaling exponents. 
Because of the limitation of disk geometry,  we put the fractional quantum Hall (FQH) state on the surface of a cylinder which has a larger tunable tunneling distance 
than that on disk by varying the aspect ratio $\gamma$. We analyze the scaling behavior of the quasiholes, especially the non-Abelian quasiholes in 
 the Read-Rezayi ${\mathbb Z}_k$ parafermion states. We aim to address the existance of the anomalous correction of the scaling parameter in the long tunneling distance.
\end{abstract}
\maketitle

\section{Introduction}
According to the experimentally realizable phases that support topological objects, 
the fractional quantum Hall (FQH) effect, since its discovery~\cite{Tsui}, has appealed to tremendous theoretical and experimental attractions and 
achieved us a collection of methods to study the strongly correlated electron systems.  The quasiparticle excitations in the FQH liquids can have 
fractional charges and obey fractional statistics~\cite{Wilczek, Halperin, Arovas}.
Within a serious of filling factors of FQH states, some of them may support more exotic excitations with non-Abelian statistics, which have potential applications
in the topological protected quantum computation~\cite{Kitaev, Preskill, Freedman, RMPNayak}. The FQH with an even denominator on the first Landau level 
at filling factor $\nu = 5/2$~\cite{Willett} is the most studied state that belongs to the family of non-Abelian FQH states.  Since the seminal work of Moore and Read~\cite{Moore},
a connection between the wavefunction of the FQH state and the conformal field theory (CFT) has been established. Thereafter, a series of non-Abelian FQH states
have been proposed which are described by the $SU(2)_k$ topological quantum field theory~\cite{Read}. The index $k$ describes the clustering properties 
in the model wavefunction and has a connection with its filling factor $\nu = \frac{k}{kM+2}$. They are addressed Read-Rezayi ${\mathbb Z}_k$ quantum Hall 
states since the corresponding wavefunctions can be calculated from the correlation functions in the ${\mathbb Z}_k$ parafermionic conformal field theory~\cite{Fateev}.
Notablely, it was found~\cite{Read} that the ${\mathbb Z}_k$ FQH states are the exact ground state of certain Hamiltonian with $k+1$-body interaction for all integers $k \geq 1$. 
Further more, Bernevig et.al. ~\cite{bernevig08, bernevig08a, bernevig09} recently found that the homogeneous polynomial part of the ${\mathbb Z}_k$ FQH wavefunction can be obtained recursively from the Jack polynomials
which is one of the polynomial solutions for Calogero-Sutherland Hamiltonian~\cite{Feigin02}.   The ${\mathbb Z}_k$ FQH wavefunctions in this language are labelled by
 a negative parameter $\alpha$ and a root configuration (or partition). Because of its computation advantages, we thereafter
 use the Jacks to produce the model wavefunctions for the ground state and their quasiparticle excitations in the following calculation.

The measurement of the transport properties of the quasiparticles propagating along the edge of the FQH states is crucial for identification of the topological nature of the 
systems. As standard practice in the noise and interference experiments~\cite{chamon97, Fradkin, Sarma, stern06, bonderson06, Rosenow, Bishara, Bonderson, Fidkowski, Bonderson08, Ardonne},
quantum point contacts are introduced to allow quasiparticles propagating on one edge to tunnel to another.  This motivated us to study the quasiparticle tunneling amplitudes
in FQH liquids in the disk geometry~\cite{Chen, Hu2011, Hu2012a}.  On the disk, we considered a tunneling potential along a specific direction
 $V_{\text{tunnel}} = V_t\delta(\theta)$.  The tunneling amplitudes exhibit interesting scaling behavior, whose exponent is related to the conformal dimension and the charge
 of the tunneling quasiparticles. Specificaly, from an effective field theory analysis, we found~\cite{Hu2011} that the bare quasiparticle tunneling matrix element satisfies a scaling function
 \begin{eqnarray} \label{scaling}
  \Gamma_a = \langle0|H_T|\Psi_a^{qh}\rangle \propto N^{1-2\Delta_a} K_a(d) = N^\alpha K_a(d),
 \end{eqnarray}
where $|0\rangle$ and $|\Psi_a^{qh}\rangle$ are the ground state and quasihole wavefunction respectively and $\Delta_a$ is the conformal dimension of the quasiparticle operators.
$H_T = \sum_a t_a[\Psi_{a, 1}^\dagger(0) \Psi_{a,2}(0) + h.c. ]$ is the edge-edge tunneling Hamiltonian.
The function $K_a(d)$ reveals the dependence on the tunneling distance $d$.  
We analyzed the tunneling amplitude of the Abelian quasiparticle, such as the $e/3$ and $2e/3$ quasiholes
in the Laughlin state at $\nu = 1/3$ and $e/2$ quasihole in the Moore-Read state at $\nu = 5/2$. As shown in Table 1 of Ref.~\onlinecite{Hu2011},
an excellent agreement with the above relation $\alpha = 1 - 2 \Delta_a$ was found.  However, for non-Abelian excitations, such as $e/4$ quasihole excitation in the Moore-Read state, 
it is likely that there is a correction on the scaling parameter, i.e., 
\begin{eqnarray}\label{modifiedalpha}
 \alpha^{e/(k+2)} = 1 - 2\Delta_a - \frac{k-1}{2k}.
\end{eqnarray}
Incidentally, the anomalous term can be written as $-(k+2)\Delta_n$ where $\Delta_n$ is the conformal dimension for the neutral part of the non-Abelian quasihole operator. We therefore,
in the case of disk geometry, had an argument that this anomalous term may origin from the unconstrained tunneling of the neutral parafermions.  However,  because of the curvature the disk geometry,
for a finite size system, the tunneling distance is limited by the radius of the system which is proportional to $\sqrt{N}$.  A natural way to overcome this problem is putting the electrons
on cylinder or torus. Comparing to the disk, the cylinder geometry has advantages that there is no curvature difference between
Landau orbitals and the edge-edge distance is linearly proportional to the system size $N$.  Moreover, the edge-edge distance can be easily tuned via varying the aspect ratio, by comparison,
in a longer range. Our recent work on the quasihole tunneling and entanglement entropy in Laughlin state on cylinder~\cite{Qi} found that a critial length scale of the edge-edge distance exists.
It can be explained as a threshold value that the two edges of the cylinder can be treated independently, or the effects of the edge-edge interaction can only be neglected while $d > L_x^c$ 
where $L_x^c \sim 5 l_B$. Therefore if we  want to consider the scaling behavior of the quasiparticle tunneling 
between two independent edges with a longer tunneling distance, the cylinder geometry is a better choice than disk.  In this paper, we study the physical properties 
of the FQH liquids and reconsider the quasiparticle tunneling amplitudes scaling on cylinder, especially focus on the scaling of the non-Abelian quasiparticles tunneling. 
We aim to address the question of the existence of the anomalous term in the scaling parameter for non-Abelian quasiholes in the whole region of tunneling distance.  

The remainder of this paper is organized as follows. Section \uppercase\expandafter{\romannumeral2} gives a brief review of the mode and previous results in the disk geometry . 
The model of  a quasihole tunneling on cylinder is introduced in section \uppercase\expandafter{\romannumeral3}.
Section \uppercase\expandafter{\romannumeral4} is devoted to the scaling analysis for the Abelian and non-Abelian quasiholes in ${\mathbb Z}_2$ state. 
In section \uppercase\expandafter{\romannumeral5} we focus on the scaling analysis in other ${\mathbb Z}_k$ states such as the $k = 3$ and $k = 4$ cases. 
Section \uppercase\expandafter{\romannumeral6} provides a conclusion and  discussion of the paper.

\section{Model and previous results}
In the disk geometry~\cite{Chen}, we considered a single-particle tunneling potential $V_\text{tunnel} = V_t \delta(\theta)$, which breaks the rotational symmetry.  It defines a tunneling path for FQH quasiparticles
under the gate influence at a quantum point contact.  The tunneling amplitude for a quasihole to the FQH droplet edge is  $\Gamma_{qh} = \langle \Psi_{qh}| \sum_i V_{\text{tunnel}}(\theta_i)|\Psi_0\rangle$. 
The most efficient way of getting these wavefunction is by the help of Jack polynomial approachment~\cite{bernevig08, bernevig08a, bernevig09}. 
Generally, Jacks belongs to a family of symmetric multivariate polynomials of the complex particle coordinates. Potentially,
they can be the bosonic version of FQH wavefunction (appending the ubiquitous Gaussian factor $e^{-\sum_i |z_i|^2/4}$ on disk), or fermionic version with an extra Vandermonde determinant $\prod_{i<j} (z_i - z_j)$. 
A Jack $J_\lambda^\alpha(z_1, z_2, \cdots, z_N)$ is parameterized by a negative rational number $\alpha$, which is related to the clustering properties of the wavefunction, and a root configuration $\lambda$,
which satisfies a generalized Pauli exclusion principle and from which one can derive a set of monomials that form a basis for the wavefunction. For the Read-Rezayi ${\mathbb Z}_k$ parafermion state,
the $\alpha = - k - 1$ and the corresponding root configuration can be expressed by a binary format ``$1^k0^21^k0^2\cdots$''  in the occupation representation of Landau orbitals.
Taking the Moore-Read state with $k = 2$ as an example, the root configuration for ground state is ``$11001100\cdots$'' 
where the leftmost orbital represents the innermost Landau orbital with the symmetric gauge. In order to vary the tunneling distance, a large number of Abelian quasihole with charge $e/2$
were inserted at the center, namely a bunch of zeros are attached in front of the root configurations. Therefore, the tunneling distance for a $N$-electron ${\mathbb Z}_k$ parafermion state is 
\begin{eqnarray}
 d(n, N, k) / l_B = \sqrt{2n + \frac{2N(k+2)}{k} - 4} - \sqrt{2n},
\end{eqnarray}
in which $n$ is  the number of quasiholes inserted at the center.  The system evolves from disk to  annulus and finally to  a ring shape while increasing $n$.
The tunneling distance $d \rightarrow 0$ in the one dimensional ring limit while $n\rightarrow \infty$. In this limit, we had a conjecture that the tunneling amplitudes 
for  Abelian quasihole with charge $\frac{ke}{k+2}$ for ${\mathbb Z}_k$ parafermion state is 
\begin{eqnarray} \label{Abelianformula}
 2 \pi \Gamma_k^{ke/(k+2)}(N) = \frac{N}{k+2} B\left(\frac{N}{k}, \frac{k}{k+2} \right),
\end{eqnarray}
where we introduce the $\beta$ function $B(x, y) = \Gamma(x) \Gamma(y) / \Gamma(x + y)$.   Unfortunately,  we did not find a universal formula for the non-Abelian
quasihole tunneling amplitudes for ${\mathbb Z}_k$ FQH states except for the $e/4$ quasihole in Moore-Read state at $k = 2$: 
\begin{equation} \label{quarterformula}
2 \pi \Gamma^{e/4} (N) = {N/2 \over 4}
  \sqrt{ B \left ({N \over 2}, {1 \over 2} + {\sqrt{3} \over 4} \right )
    B \left ({N \over 2}, {1 \over 2} - {\sqrt{3} \over 4} \right )}.
\end{equation}
\begin{figure}
 \includegraphics[width=8cm]{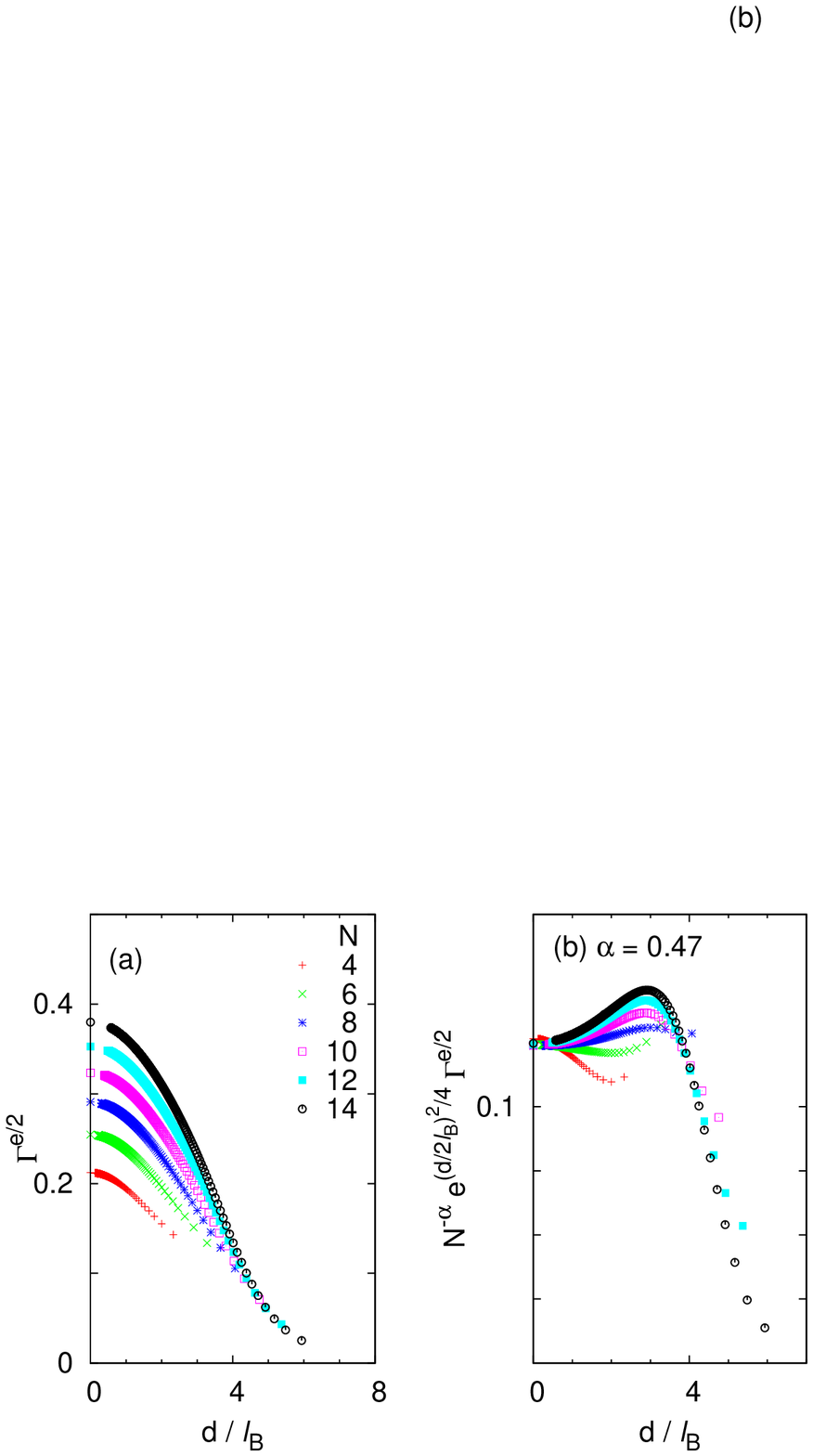}
 \caption{\label{diskhalf}The unrescaled (a) and rescaled tunneling amplitude $N^{-\alpha}e^{(d/2l_B)^2/4}\Gamma^{e/2}$ with $\alpha = 0.47$ (b) for e/2 quasihole 
 in the MR state as a function of the tunneling distance $d$ on disk.  }
\end{figure}
Before discussing the scaling behavior of the tunneling amplitudes of the ${\mathbb Z}_k$ parafermion states, we firstly write down the formula of the scaling dimension for the quasiholes.
In ${\mathbb Z}_k$ parafermionic CFT, the Abelian quasihole operator is $\psi_{qh}^{\text{Abelian}} = e^{i\phi\sqrt{k/(k+2)}}$ where $\phi$ is the charge bosonic field.
The Abelian quasihole has charge $\frac{ke}{k+2}$ and scaling dimension $\Delta_{ke/(k+2)}^{\text{Abelian}} = \frac{k}{2(k+2)}$.  
On the other hand, the operator for the smallest charged non-Abelian quasihole is $\psi_{qh}^{\text{non-Abelian}} = \sigma_1 e^{i\phi/\sqrt{k(k+2)}}$ in which the $\sigma_1$ is the 
neutral spin fields which has scaling dimension $\Delta_{\sigma_1} = \Delta_n = \frac{k-1}{2k(k+2)}$. Therefore, the smallest charged non-Abelian quasihole has charge $\frac{e}{k+2}$
and scaling dimension $\Delta_{e/(k+2)}^{\text{non-Abelian}} = \Delta_{\sigma_1} + \frac{1}{2k(k+2)} = \frac{1}{2(k+2)}$.
Eq.(\ref{Abelianformula}) gives the asymptotic scaling behavior for Abelian quasiholes in the ring limit $\Gamma_k^{ke/(k+2)}(N) \sim N^{1-k/(k+2)} = N^{1-2\Delta_{ke/(k+2)}}$.
Combining with the long distance behavior with a gaussian decay, we conjectured a scaling function as
\begin{eqnarray}\label{scalingconj}
 \Gamma^q(N, d) = \Gamma_0 N^{\alpha^q} e^{-(qd/2el_B)^2} ,
\end{eqnarray}
where $\alpha^q = 1-2\Delta^q$ is the scaling parameter.
Fig.~\ref{diskhalf} shows the bare and rescaled tunneling amplitudes for the Abelian $e/2$ quasihole as a function of $d$ on a disk.
With the scaling parameter $\alpha^{e/2} = 0.47 \simeq 1-2\Delta^{e/2} = 0.5$, the data for systems from $4-14$ electrons collapse to the same value 
while $d \rightarrow 0$ which are in agreement with the scaling conjecture of Eq.(\ref{scalingconj}).

\begin{figure}[H]
 \includegraphics[width=8cm]{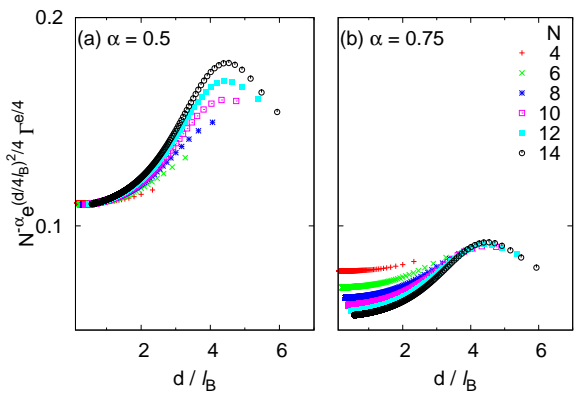}
 \caption{\label{diskquarter}The rescaled tunneling amplitude $N^{-\alpha}e^{(d/4l_B)^2/4}\Gamma^{e/4}$ for e/4 quasihole in the ${\mathbb Z}_2$
 parafermion, or Moore-Read state as a function of the tunneling distance $d$ on disk.  }
\end{figure}

The more interesting is for the tunneling amplitudes of the non-Abelian quasihole. In Fig.\ref{diskquarter}, we present two rescaled data for $e/4$ quasihole in 
Moore-Read state. It is shown that the data from different system sizes in the ring limit collapse well at $\alpha = 0.5$ instead of at $\alpha = 0.75 = 1-2\Delta^{e/4}$. 
Since the correction of the scaling parameter for ${\mathbb Z}_k$ with $k=2, 3, 4, 5$ can be approximated by $-(k+2)\Delta_{\sigma_1}$, we speculated~\cite{Hu2011}
that the anomalous scaling behavior for the non-Abelian quasihole may origin from the effect of non-independent edges for neutral component, or the edge-edge interactions while in small $d$.  
On the other hand, if we look carefully at the Fig.\ref{diskquarter}(b), for the range $d > 5l_B$, although few data points limited by
the geometry and system size, there is still a hint that the scaling behavior works well in large $d$ regime without the neutral part correction. Therefore, a natural question
is that whether the scaling conjection of Eq. (\ref{scalingconj}) works for large tunneling distance? With this motivation, we reconsider the scaling behavior of the
tunneling amplitudes for ${\mathbb Z}_k$ parafermion states on cylinder in the following section.

\section{Quasihole tunneling on cylinder}
 \begin{figure}[H]
 \includegraphics[width=8cm]{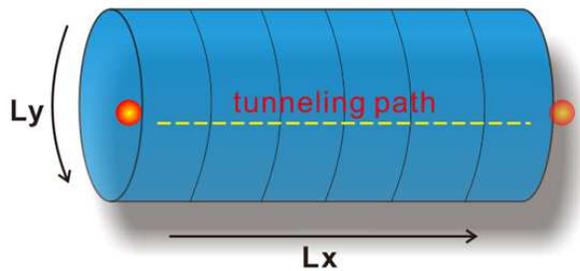}
 \caption{\label{cylinder}The sketch of a cylinder we used to make the quasihole tunnel from the left edge to the right under a tunneling potential $V_{\text{tunnel}} = V_t\delta(y)$. 
 $L_y$ is the circumference of the edge and $L_x$ is the length of the finite cylinder. }
 \end{figure}
As modelled in Fig.~\ref{cylinder}, for a cylinder with circumference $L_y$ in $y$ direction, the lowest Landau level wave function for single electron in a magnetic field with Landau gauge is:
\begin{equation}                                                                                                                                  
\psi_{j}(\vec{r})=\frac{1}{\sqrt{\pi^{1/2}L_y}}e^{ik_{y}y}e^{-\frac{1}{2}(x+k_{y})^{2}},
\end{equation}    
in which $ k_{y}=\frac{2\pi}{L_y}j $, $ j=0,\pm1,\pm2\cdots $  are the equilibrium positions for each Landau orbital. Here the magnetic length $l_B = \sqrt{\hbar c/eB}$ 
has been set to one. The magnetic field is perpendicular to the cylinder surface and the number of orbits $N_{orb}$ equals to the number of magnetic flux 
quantum penetrating from the surface. As each state or each orbit occupies a constant area $2\pi l_B^2$, the total area  is 
$ A = 2 \pi l_B^2 N_{orb} $ for a finite size system. The aspect ratio is defined as $\gamma = L_y/L_x$. 

To study the quasiparticle tunneling on cylinder, we use a simple delta tunneling potential $V_{\text{tunnel}}=V_t\delta(y)$ for a single particle similar to 
the disk geometry~\cite{Chen, Hu2011}. This potential allows a quasiparticle to tunnel along the $x$-direction while the y-direction doesn't have translational 
symmetry.   The matrix element 
$ \langle k \vert V_{\text{tunnel}} \vert m \rangle $  describes a particle tunneling from one single particle state
$\vert m \rangle $ to another state $\vert k \rangle$. If we set $V_t = 1$ for convenience, it is easy to get the matrix element form as follows:
\begin{eqnarray}                                                                                                                                  
v_p(k,m)&=&\langle{k}\vert{V_{\text{tunnel}}}\vert{m}\rangle = e^{-\frac{(\frac{2\pi}{L_y}m-\frac{2\pi}{L_y}k)^2}{4}} \nonumber \\
&=&e^{-\frac{(d/l_B)^2}{4}} ,
\end{eqnarray}  
where $d$ is the distance between two single states. 
In many-body case, the tunneling operator is just the summation of the single particle tunneling potential $H_T = V_t\sum_i\delta(y_i)$. Then the tunneling amplitude is obtained by 
calculating 
\begin{eqnarray} \label{tunnelingformula}
\Gamma &=&\langle \psi_{qh} \vert \tau \vert \psi_{0} \rangle \nonumber\\                                                                                       
&=& \sum_i\langle k_1k_2\cdots k_n\vert \delta (y_i)C_{k}^+C_m \vert m_1m_2\cdots m_n \rangle ,
\end{eqnarray} 
where $\vert k_1k_2\cdots k_n \rangle \in \psi_{qh}$ and $\vert m_1m_2\cdots m_n\rangle \in \psi_0 $.
The matrix elements have non-zero components coming from the identical sets $\vert k_1k_2\cdots k_n \rangle $ and $\vert m_1m_2\cdots m_n\rangle$ except a single pair $m'$ and $k'$ which 
have constant value of momentum difference. Taking Moore-Read state as an example, the $e/2$ case need the $k'-m' = N$ while the $e/4$ case need the $k'-m' = N / 2$ where 
$N$ is the number of electrons.  Thus, we easily get 
$v_p^{e/2} = e^{-\frac{\pi^2}{L_y^2}N^2}$ and $v_p^{e/4} = e^{-\frac{\pi^2}{4L_y^2}N^2}$, or the charge of the quasihole should appear in the Gaussian factor.

\section{Scaling analysis for the ${\mathbb Z}_2$ state}
In this section, we systematically study the scaling behavior of the tunneling amplitude in the case of ${\mathbb Z}_2$, or Moore-Read state on a cylinder.
In the language of the Jack polynomial description, the Moore-Read state and its quasihole state are labelled by root configurations 
$\vert \psi_{\text{MR}} \rangle = \vert 11001100\cdots110011 \rangle$, $\vert \psi_{qh}^{e/2}\rangle =\vert 011001100\cdots110011 \rangle$
and $ \vert \psi_{qh}^{e/4}\rangle =\vert 10101010\cdots 0101 \rangle $ respectively.  Here the extra zero in the $\vert \psi_{qh}^{e/2}\rangle$ on the left means
a flux quantum, or an Abelian $e/2$ quasihole is created on the left edge of the cylinder.  The pattern in the $\vert \psi_{qh}^{e/4}\rangle$ means there are two $e/4$ quasiholes,
one on each edge of the cylinder since the non-Abelian Majorana fermion modes are embedded in the $e/4$ quasihole excitation~\cite{Moore} which must appear in pair. 
As was introduced in above section, the tunneling path of the quasihole in this case is the length of the finite cylinder in $x$-direction, namely $d = L_x = \sqrt{2\pi N_{orb}/\gamma}$.
In this case, we have a parameter $\gamma$ that can smoothly tune the tunneling distance from zero to infinity. As a comparison, in the disk geometry,  
the tunneling distance has a upper limit which is the radius of the system $R = \sqrt{2N_{orb}}$ and $d \sim N_{orb}/\sqrt{2N_{qh}}$ for $N_{qh} \gg N_{orb}$, 
thus the disk geometry is not suitable to large $d$ physics. Interestingly, in the limit of  $L_x \rightarrow \infty$ or $\gamma \rightarrow 0$, the two adjacent Landau 
orbitals have practically zero overlap and thus the Hamiltonian is dominated by the electrostatic repulsion. This thin cylinder limit has  ground state which is called 
a charge density wave state, or Tao-Thouless state~\cite{tao1983, thouless84} on torus with occupation pattern $1001001001\cdots$. However previous studies
in this limit~\cite{Seidel, Bergholtzprb08, Bergholtzprb08a} shows that the topological properties of the FQH state does not change as varying $\gamma$, or there is no phase transition
while varying the aspect ratio.  Therefore, we feel more comfortable to say that the results calculated from Eq.(\ref{tunnelingformula}) in the whole range of $L_x$  
are really the tunneling amplitudes for FQH quasiholes.
\begin{figure}
\centering
\includegraphics[width=9cm,height=6cm]{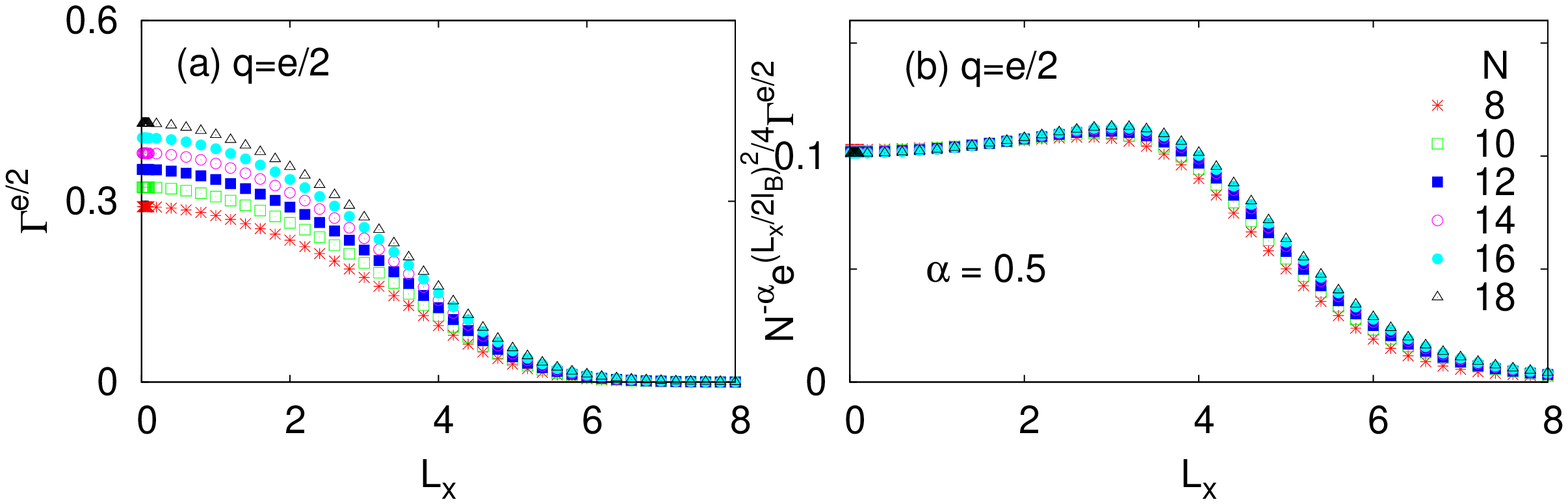}
\caption{\label{halfcylinder}The bare (a) and rescaled tunneling amplitude $N^{-\alpha}e^{(L_x/2l_B)^2/4}\Gamma^{e/2}$ with $\alpha = 0.5$ (b) for e/2 quasihole 
 in the Moore-Read state as a function of the tunneling distance $L_x$ on cylinder. The system size ranges from 8 to 18 electrons.}
\end{figure}
Since the Landau wavefunctions both on the disk and cylinder have the similar feature to Gaussian wave, following the work on disk~\cite{Hu2011}, we believe 
that the quasiparticle tunneling amplitude on cylinder has the same scaling behavior as that on disk in Eq.(\ref{scalingconj}).
Fig.~\ref{halfcylinder}(a) shows that the tunneling amplitudes for $e/2$ quasihole as a function of the tunneling distance $L_x$.  By the help of Jacks, the largest system size
we have reached is $18$ electrons. The data is the same as that on disk in Fig.~\ref{diskhalf}(a) except that there are more data for large $L_x$ on cylinder. In the limit $L_x \rightarrow 0$,
or the CFT limit, with the same argument as that on disk, the value of the tunneling amplitude are exactly the same as that in Eq.~(\ref{Abelianformula}).  As $L_x$ increasing, the tunneling amplitudes 
dramatically drop to zero which dominated by the gaussian factor in Eq.(\ref{scalingconj}).

In Fig.~\ref{halfcylinder}(b),  we  plot the rescaled data $\tilde{\Gamma} = N^{-\alpha} e^{(qL_x/2el_B)^2}\Gamma^{q}$ as a function of the tunneling distance.  While $\alpha = 1  - 2\Delta^{e/2} = 0.5$, the 
scaling behavior is obviously better than that on disk as shown in Fig.~\ref{diskhalf}(b). The data from $8 - 18$ electrons not only in the CFT limit $L_x \rightarrow 0$, but also 
in the large $L_x$ region collapse onto each other.  Therefore we see that the scaling function still works, for charged $e/2$ Abelian quasihole,  in the region far away from the CFT limit.
The next question is whether it is workable for the non-Abelian quasihole tunneling amplitude, and whether the anomalous term in the scaling parameter still exist?  

\begin{figure}[H]
\centering
\includegraphics[width=9cm,height=6cm]{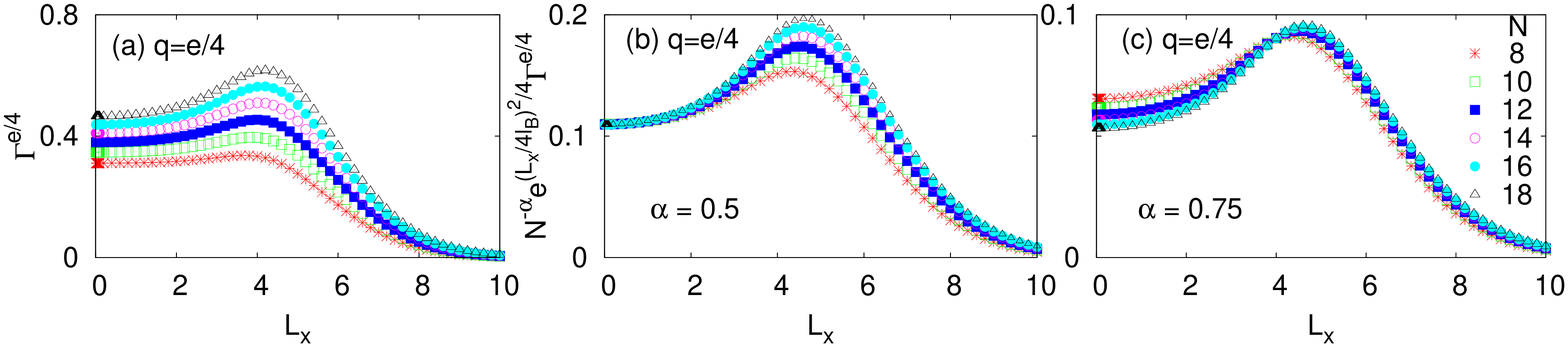}
\caption{\label{quartercylinder}The unrescaled (a) and rescaled tunneling amplitude $N^{-\alpha}e^{(L_x/4l_B)^2/4}\Gamma^{e/2}$ with $\alpha = 0.5$ (b) $\alpha = 0.75$ (c) for e/4 quasihole 
 in the MR state as a function of the tunneling distance $L_x$ on cylinder. }
\end{figure}

In Fig.~\ref{quartercylinder}(a), we plot the tunneling amplitude for $e/4$ quasihole as a function of $L_x$ for systems ranges from 8 to 18 electrons.
The values in the CFT limit are still consistent to the Eq.(\ref{quarterformula}).  Comparing with the
Abelian case as  shown in Fig.~\ref{halfcylinder}(a), the bare tunneling amplitudes for non-Abelian $e/4$ quasihole do not decay monotonically as increasing tunneling distance.
There is a bump around $L_x \simeq L_x^c \simeq 5l_B$ which is consistent to the critial threshold value at which the interaction between two edges of FQH states can not be neglected~\cite{Qi}.
The more interesting is that, as shown in  Fig.~\ref{halfcylinder}(b), while the tunneling amplitudes are rescaled by Eq.(\ref{scalingconj}) with a modified scaling parameter $\alpha^{e/4} = 0.5$ 
as expressed in Eq.(\ref{modifiedalpha}),
the data approaching to the CFT limit still scales very well. This result has consistency with that in the disk geometry. However, with this scaling parameter,  the 
tunneling amplitudes for large $L_x$ do not collapse onto each other.  Since the two edges of the  cylinder in this case are getting more and more independent while increasing the edge-edge
distance, it motivates us to look the scaling behavior without the correction term in Eq.(\ref{modifiedalpha}).  The results are shown in Fig.~\ref{quartercylinder}(c). Here we set the scaling
parameter $\alpha = 1 - 2\Delta_a = 0.75$. It is shown that the rescaled data for all systems scales very well in the range $L_x > L_x^c$
and obviously, the scaling behavior in the CFT limit in this case is broken in this case. 

As a conclusion for this section,  we find that in the cylinder geometry, the tunneling amplitude for the Abelian $e/2$ quasihole in the Moore-Read state obeys the scaling
behavior of Eq.(\ref{scalingconj}) with $\alpha^q = 1 - 2\Delta_a$ both in the region with short tunneling distance (CFT limit) and in the region of long tunneling distance. 
However, for the non-Abelian $e/4$ quasihole, the above scaling parameter is only workable for $L_x > L_x^c$ where the $L_x^c$ is the critical distance, above which the two edges
of the cylinder can be treated as two independent ones.  As the previous study on disk, the data for the tunneling amplitudes in $L_x < L_x^c$ obey
the similar scaling function with a non-trivial modification on the scaling parameter $\alpha$ as expressed in Eq.~(\ref{modifiedalpha}).

\section{Scaling analysis in other ${\mathbb Z}_k$ states}
To furtherly confirm that our conslusions for the case of $k = 2$ are suitable for all the ${\mathbb Z}_k$ parafermion FQH states. In this section, we look at the scaling behavior
and the related scaling parameter in the cases of $k = 3$ and $k = 4$.  First of all, since we need to consider the relation between the scaling parameter $\alpha$ and the scaling
dimension of the quasiholes in FQH states, we list the quasihole charge, scaling dimension and its corresponding scaling parameters
for Abelian and non-Abelian quasiholes in the following two tables.
\begin{table}[H]
  \caption{\label{tbl:abelianRR}
    The scaling exponents for charge $ke/(k+2)$ Abelian quasiholes tunneling amplitudes scaling in the Read-Rezayi states. $\Delta_{qh}$ is the scaling dimension for quasihole.}
\begin{center}
\begin{tabular}{c c c c}
\hline \hline
\hspace{0.5cm} $k$ \hspace{0.25cm}
& \hspace{0.25cm} Q \hspace{0.25cm}
& \hspace{0.25cm} $\Delta_{qh} = \frac{k}{2(k+2)}$ \hspace{0.25cm}
& \hspace{0.25cm} $\alpha = 1 - 2\Delta_{qh} $ \hspace{0.25cm} \\
\hline
3 & $3e/5$ & 3/10 & 2/5 \\
4 & $4e/6$ & 1/3 & 1/3 \\
\hline
\end{tabular}
\end{center}
\end{table}
\begin{table}[H]
  \caption{\label{tbl:abelianRR}
    The scaling exponents for charge $e/(k+2)$ non-Abelian quasiholes tunneling amplitudes scaling in the Read-Rezayi states. $\Delta_{qh} = \Delta_{c} + \Delta{n} $ is 
    the scaling dimension for non-Abelian quasihole, where $\Delta_{c} = \frac{1}{2k(k+2)}$ and $\Delta_{n} = \frac{k-1}{2k(k+2)}$. }
\begin{center}
\begin{tabular}{c c c c c c }
\hline \hline
\hspace{0.2cm} $k$ \hspace{0.1cm}
& \hspace{0.1cm} Q \hspace{0.1cm}
& \hspace{0.1cm} $\Delta{c}$ \hspace{0.1cm}
& \hspace{0.1cm} $\Delta_{n}$ \hspace{0.1cm}
& \hspace{0.1cm} $ 1 - 2\Delta_{qh} - \frac{k-1}{2k} $ \hspace{0.1cm} 
& \hspace{0.1cm} $ 1 - 2\Delta_{qh}  $ \hspace{0.1cm}\\
\hline
3 & $e/5$ & 1/30 & 1/15 & 7/15(0.4667) & 4/5(0.8) \\
4 & $e/6$ & 1/48 & 1/16 & 11/24(0.45833) & 5/6(0.8333)\\
\hline
\end{tabular}
\end{center}
\end{table}
\begin{figure}
\includegraphics[width=9cm,height=6cm]{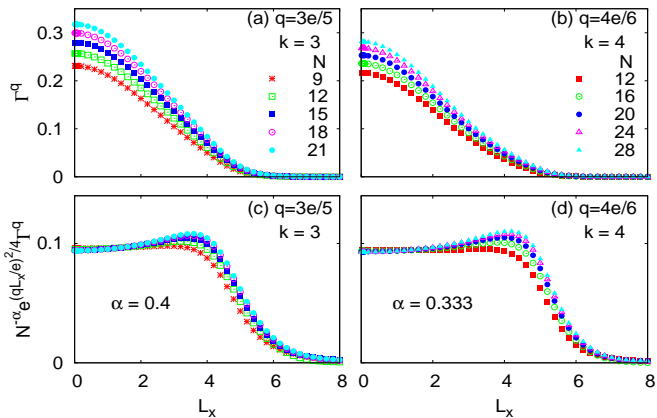}
\caption{\label{k34abelian}The unrescaled tunneling amplitude $\Gamma^q$ (a)(b) and rescaled tunneling amplitudes $N^{-\alpha}e^{(qL_x/e)^2/4}\Gamma^q$ (c)(d) for charge $ke/(k+2)$ Abelian quasihole 
 in the ${\mathbb Z}_k$ states as a function of the tunneling distance $L_x$ on cylinder. }
\end{figure}

Due to the limitation of storing the basis of the wavefunction,  the system sizes for ${\mathbb Z}_3$ and ${\mathbb Z}_4$ parafermion states we considered are up to 21 and 
28 electrons respectively. As shown in Fig.\ref{k34abelian}, like in the ${\mathbb Z}_2$  case,   the tunneling amplitudes for Abelian quasiholes in the ${\mathbb Z}_3$ and ${\mathbb Z}_4$ states monotonically decay
exponentially as increasing the tunneling distance.  After being rescaled by Eq.~(\ref{scalingconj}), all the data locate on the same curve except for that of the smallest system size.
The  scaling parameters $\alpha^q$ we used in the plot are exact the expect values $\alpha^q = 1 - 2\Delta_a$.  Here we should notice that the finite size effect becomes
more and more significant with increasing the $k$ index.

\begin{figure}
\includegraphics[width=9cm,height=6cm]{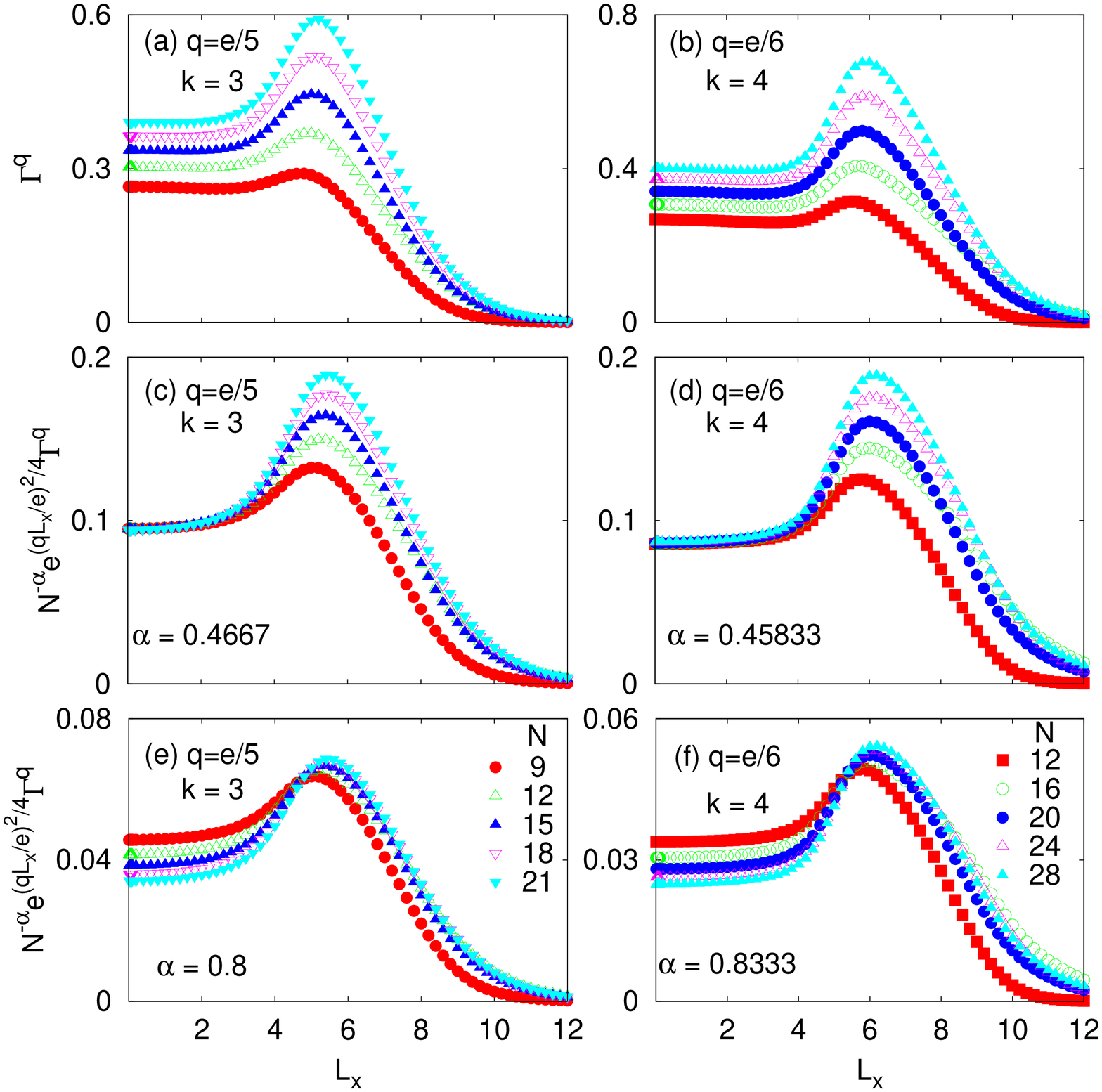}
\caption{\label{k34nonabelian}The unrescaled tunneling amplitude $\Gamma^q$ (a)(b) and rescaled tunneling amplitudes $N^{-\alpha}e^{(qL_x/e)^2/4}\Gamma^q$ (c)-(f) for charge $e/(k+2)$ non-Abelian quasihole 
 in the ${\mathbb Z}_k$ states as a function of the tunneling distance $L_x$ on cylinder.}
\end{figure}

Again, for the smallest charged non-Abelian quasihole in ${\mathbb Z}_3$ and ${\mathbb Z}_4$ states, as shown in Fig.\ref{k34nonabelian}(a) and (b),
there are still bump shapes in the tunneling amplitudes as varying $L_x$. In the CFT limit, although we do not have an analytical expression for the non-Abelian quasihole
tunneling amplitude as that for the Moore-Read state in Eq.~(\ref{quarterformula}), we believe that they still obey the similar scaling behavior.
Therefore, we firstly replot the rescaled data with the scaling parameter modified by $-\frac{k-1}{2k}$ which is shown in Fig.\ref{k34nonabelian}(c) and (d).  As expected, 
the data from different systems collapse together near the CFT limit, or in the edge-edge interacting case. This again verifies that the interaction of the two edges of the cylinder
brings the anomalous correction to the scaling parameter.  On the other hand, if this anomalous term is neglected, as shown in Fig.\ref{k34nonabelian}(e) and (f),
the data for the long distance tunneling scales better than that with the correction, although there are strong finite size effects.
\section{Summary and discussion}

In this work, we systematically study the scaling behavior of the quasihole tunneling amplitudes for the ${\mathbb Z}_k$ parafermion states on the cylinder geometry.
Comparing with the disk geometry we previously studied, the Landau orbitals on cylinder do not have curvature difference and the tunneling distance can be smoothly
tuned from zero to infinity via varying the aspect ratio.  While the length of the finite cylinder $L_x$ decreasing from a thin cylinder limit to the CFT limit, there is a critical length scale
 $L_x^c \simeq 5l_B$ at which the two independent edges of the cylinder become interacting with each other.  Therefore,  the length scale of the tunneling distance is automatically
 separated into two regions.   Our calculations reveals that the scaling behaviors for the overall region are in good agreement with the scaling conjecture
in Eq. (\ref{scalingconj}).  For the Abelian quasihole with charge $\frac{ke}{k+2}$ in all ${\mathbb Z}_k$ states,  by using the scaling parameter $\alpha^{qh} = 1 - 2\Delta^{qh}$
as expected from the analysis of the effective field theory, the data in two regions collapses onto each other very well. 
However, the things get more complicated in the non-Abelian case. When $L_x > L_x^c$, or with two independent edges,
the scaling parameter $\alpha$ is the same as that in the Abelian case instead of substituting the quantum dimension $\Delta^{qh}$ by the one for non-Abelian quasiholes.
And in the case of $L_x < L_x^c$, or approaching to the CFT limit, similar to the result on disk geometry, the scaling parameter need a modification which is shown in Eq.~(\ref{modifiedalpha}).
Based on the following reasons:  (1) the tunneling amplitudes of the  Abelian quasiholes do not have this term in the CFT limit;
(2) this anomalous term can be rewritten as $-(k+2)\Delta_n$ where $\Delta_n$ is the conformal dimension of the neutral component;  
(3) our recent work elsewhere~\cite{NaJiang} on the density difference between the bosonic and fermionic edge states shows that the width 
of the fermionic edge states in ${\mathbb Z}_k$ states is larger than that of bosonic edge states. Or in other words, the neutral fermionic component 
in the edge state or quasihole state is more sensitive to the length scale of the edge-edge distance. We conclude that the anomalous correction term is contributed from the neutral component
of the non-Abelian quasiholes which can be treated as another feature of the non-Abelian quasiholes.
This charactistics may be detected in the realistical systems, such as in the shot noise or the
point contact interference experiments. 

This work was supported by NSFC Project No. 1127403, 11547305, Fundamental Research Funds for the Central Universities No. CQDXWL-2014-Z006.  
NJ was also supported by Chongqing Graduate Student Research Innovation Project No. CYB14033.  XW was supported by the 973 Program Project
No.  2012CB927404, NSF-China Grant No.  11174246.

\setcounter{section}{0}
\renewcommand\thesection{\Alph{section}}
\numberwithin{equation}{section}

\end{document}